\documentclass[a4paper,12pt]{article}

\usepackage{amsmath}
\usepackage{graphicx}
\usepackage{hyperref}
\usepackage{bm}

\usepackage[margin=1.5cm]{geometry}
\usepackage{multicol}

\providecommand{\grad}{\nabla}
\providecommand{\bra}[1]{\ensuremath{\left(#1\right)}}
\newcommand{\sbt}{\lambda}

\begin{document}

\title{Simulations of particle tracking in the oligociliated mouse node and implications for left-right symmetry breaking mechanics}

\author{M.T.~Gallagher\thanks{Address for correspondence: m.t.gallagher@bham.ac.uk}, T.D.~Montenegro-Johnson and D.J.~Smith}

\date{School of Mathematics, University of Birmingham, Birmingham, B15 2TT. UK.}

\maketitle

\begin{abstract}
    The concept of internal anatomical asymmetry is familiar; usually in humans the heart is on the left and the liver is on the right, however how does the developing embryo know to produce this consistent laterality? Symmetry breaking initiates with left-right asymmetric cilia-driven fluid mechanics in a small fluid-filled structure called the ventral node in mice. However the question of what converts this flow into left-right asymmetric development remains unanswered. A leading hypotheses is that flow transports morphogen containing vesicles within the node, the absorption of which results in asymmetrical gene expression. To investigate how vesicle transport might result in the situs patterns observed in wildtype and mutant experiments, we extend the open source Stokes flow package, NEAREST, to consider the hydrodynamic and Brownian motion of particles in a mouse model with flow driven by one, two, and 112 beating cilia. Three models for morphogen-containing particle released are simulated to assess their compatibility with observed results in oligociliated and wildtype mouse embryos: uniformly random release, localised cilium stress induced release, and localised release from motile cilia themselves. Only the uniformly random release model appears consistent with the data, with neither localised-release model resulting in significant transport in the oligociliated embryo.
\end{abstract}

\section{Introduction}

Despite external appearances, the vertebrate body plan normally exhibits a predictable left-right asymmetry in organ placement and gut looping -- in mammals typically the heart is on the left and the liver on the right. The origin of this organised symmetry-breaking -- and in particular why the heart is usually on the left and not the right -- is only partially understood. The role of motile cilia was initially suggested by Afzelius's 1976 study of the sperm flagellum in men with Kartagener's syndrome (the combination of situs inversus -- the lateral transposition of the internal organs, respiratory problems and male infertility) \cite{afzelius1976,berdon2004a,berdon2004b}. The definitive picture only emerged in the 1990s, with the discovery of a ciliated node structure in early mouse embryo development \cite{sulik1994}, and subsequent confirmation of the rotational motility of the cilia and the necessary and sufficient roles of cilia-driven fluid dynamics in establishing left-right asymmetry \cite{nonaka1998,okada1999,nonaka2002}.

\begin{figure*}
    \centering
	\includegraphics[width=\textwidth]{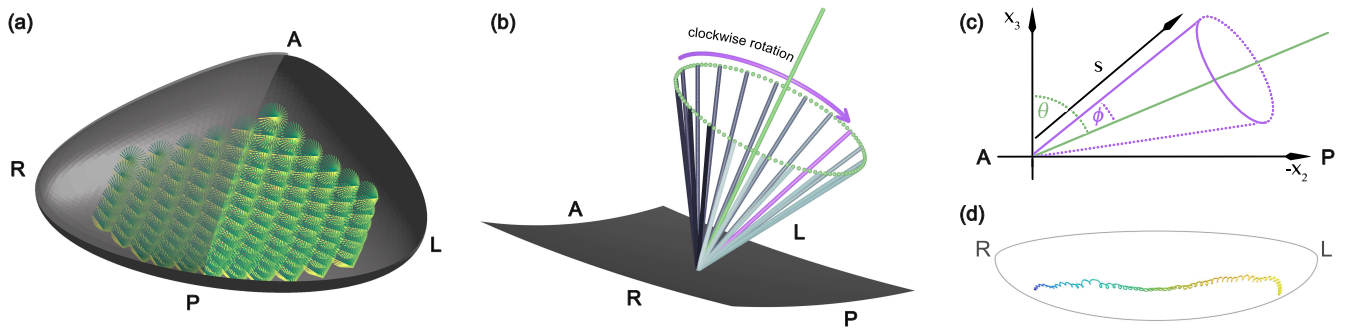}
	\caption{Sketch of (a) the geometry of the euciliated, \(112\) cilia mouse node with Reichert's membrane cut in half to show the whirling cilia, (b) schematic of an individual beating cilium shown above a section of the node, (c) sketch showing the posterior tilt \(\theta\) and semi-cone angle \(\phi\) for a model cilium, and (d) the characteristic `loopy-drift' of a particle moving through the node. Note that right and left directions are transposed as is conventional in images of this system.}
	\label{fig:sketch}
\end{figure*}

Subsequent work predicted \cite{cartwright2004}, observed \cite{nonaka2005,okada2005mechanism} and explained \cite{brokaw2005,smith2007,smith2008}, the mechanical role of posterior tilt and viscous interaction with the epithelium in converting chiral information (clockwise rotation viewed tip-to-base) to lateral information (leftward transport of particles). The physical picture is shown in figure~\ref{fig:sketch}. Whirling cilia, positioned towards the posterior of the cell body \cite{hashimoto2010}, drive stronger fluid flow during the leftward part of their stroke compared to the rightward part. This process causes a `loopy drift' of particles towards the left \cite{smith2011mathematical}. The process has been observed \emph{inter-alia} in mouse \cite{nonaka1998}, rabbit \cite{okada2005mechanism}, medakafish \cite{okada2005mechanism} and zebrafish \cite{essner2005,kramer2005}. For further details see for example the reviews  \cite{smith2019,shinohara2017,schweickert2018}.

The conversion of this leftward transport to asymmetric gene expression has however remained an enduring problem. Two conceptual models were put forward around the time of the discovery of nodal flow: the first \cite{nonaka1998} postulated that the leftward flow caused transport of diffusible morphogen and hence concentration on the left hand side of the embryo. The second model \cite{mcgrath2003} postulated that flow produces left-right asymmetric mechanical responses of immotile cilia and associated calcium signals. Neither theory has been extensively tested experimentally, and indeed theoretical complications have been raised with both. Regarding the diffusible morphogen model, the enclosed nature of the node produced by the overlying Reichert's membrane means that morphogen would need to be subject to a temporal inactivation process in order to avoid diffusion-induced uniform distribution \cite{cartwright2004}. In respect of the mechanosensing model, it is unclear how the symmetric shear stresses associated with inertialess microscale flow could be sensed asymmetrically \cite{cartwright2004}. Moreover, recent observations have raised doubt regarding whether the primary cilia found in the node are capable of producing a calcium signal at physiological shear rates \cite{delling2016}. For in-depth discussion regarding the feasibility of primary cilia as force sensors see the review \cite{ferreira2019cilium}.

A later conceptual model \cite{tanaka2005} was based on the apparent observation of `nodal vesicular parcels', membrane-enclosed vesicles containing morphogens, which were postulated to be released by microvilli from the floor of the node, transported leftward by the action of cilia, and then broken up by cilia at the left, whereupon morphogen would be absorbed. This model was attractive in that these parcels would have a lower diffusion coefficient than diffusible morphogen, and moreover would provide a mechanism for localised release. A numerical simulation study conducted by Cartwright \textit{et al.} \cite{cartwright2007} has suggested both that NVPs accumulate on the left of the node, and that the delivery of morphogens through NVP breakup ought to be through chemical interactions with the wall of the node or with certain cilia. Consistent with this work, Ferreira \textit{et al.} \cite{ferreira2017physical} simulated transport of \(2-10~\)nm molecules in Kupfer's vesicle of zebrafish, showing that anterior secretion can lead to an asymmetric absorption pattern. Despite the appeal of this theory, it has not yet been confirmed by direct observations, although it has been highlighted that vesicles can play an important role in intercellular communication (see the reviews \cite{cocucci2009,wang2018cell}).

A further puzzle -- or perhaps a clue -- was provided by the study of Shinohara \textit{et. al} \cite{shinohara2012two}, in which \textit{Dpcd} and \textit{Rfx3} mutant mouse embryos expressing as few as two cilia were found to produce consistent situs solitus. Moreover, dramatic reduction in flow induced by raising the viscosity of the medium also did not result in disturbed symmetry-breaking. These findings in the mouse node are in stark contrast to the analogous Kupffer's vesicle structure in zebrafish, in which even relatively moderate reductions to cilia activity result in major perturbations to situs \cite{sampaio2014}. The dynamics of symmetry-breaking in this extremely deciliated system -- for which we propose the nomenclature `oligociliation' -- present an interesting topic for theoretical study. Whatever the true mechanism is in mouse, it must be robust to reductions to two -- but not one -- motile cilia. Omori \textit{et. al} \cite{omori2018simulation} predicted through imaging-generated computational modelling that particle transport in the oligociliated node is greatly reduced, being tightly restricted to the region around the cilia. It was argued that these findings tend to refute the role for a transport-based mechanism. As an alternative, it was suggested that anisotropic bending stiffness of immotile cilia might enable these organelles to act as directional sensors. Building on an earlier suggestion by \cite{okada2005mechanism}, a recent study \cite{solowiej2019} found that stresses induced by the cilium beat may be sufficient to promote vesicle release from ciliated cells, indicating that vesicles may only enter the system in regions in which transport can occur. A further possibility, raised during discussions at the Theo Murphy International Meeting associated with this volume, is that vesicles may be released from cilia themselves -- again providing a local release mechanism.

In this manuscript we are motivated to progress the study of particle deposition patterns due to one, two, and many cilia, and to compare directly the effects of localised and widespread release. Working with an idealised model of the node which extends previous work by preserving the core features (concave ciliated surface, overlying membrane) and a recently-developed computational method for the Stokes flow equations (NEAREST \cite{gallagher2018meshfree}) we simulate larger numbers of particles than previously possible (approximately \(1000\) for each case) over periods of \(10,000\) beat cycles.

A key focus of the present study will be the transport of potential morphogen particles, and the influence of their origin of release in the node. Following a brief description of the mathematical model (section 2), these conceptual models of vesicle release will be assessed via simulation of oligociliated mouse nodes (section 3). Finally the models will be compared, and their consistency with reported experimental data will be discussed (section 4).

\section{Particle transport in Stokes flow}

The dynamics of particle transport in the embryonic node is dominated by the fluid velocity generated by beating cilia, and augmented by the inclusion of Brownian noise. The small length scales associated with the geometry of the node result in a low-Reynolds number regime, where viscous forces dominate inertia, resulting in a fluid behaviour that can be described by the dimensionless Stokes flow equations, namely
\begin{equation}
    -\bm{\grad} p + \grad^2 \bm{u} = 0,\qquad \bm{\grad}\cdot\bm{u} = 0.
    \label{eqn:stokes}
\end{equation}
Here, \( p \) and \(\bm{u}\) denote the fluid pressure and velocity respectively and \(\bm{\grad}\) the gradient operator in three-dimensions. We non-dimensionalise length with respect to a characteristic cilium length of \(L = 5~\mu\)m and beat frequency \(\omega\), hence the transformation from dimensional to non-dimensional variables is given by \(\bm{u}^{*} = \omega L \bm{u}\), and \(p^* = \mu\omega p\). By placing a number of rigid cilia-like rods, rotating with fixed frequency, we can solve system \eqref{eqn:stokes} subject to the no-slip, no-penetration boundary conditions on both the walls and floor of the node \(\bra{\bm{u}\bra{\bm{x},t} = \bm{0}}\), and on the cilia \(\bra{\bm{u}\bra{\bm{x},t} = \bm{\dot{x}}}\). We discretise \eqref{eqn:stokes} over the node boundary using the meshfree regularised Stokeslet implementation NEAREST \cite{gallagher2018meshfree}, via a coarse-force discretisation \(\bm{x}^b [n]\) \(\bra{n = 1,\hdots, N^b}\), and corresponding fine-quadrature discretisation \(\bm{X}^b [q]\) \(\bra{q = 1,\hdots, Q^b}\). For the rotating cilia we discretise into segments of equal length \(\Delta s\) located at \(\bm{x}^c [m]\) \(\bra{m = 1,\hdots N^c}\). For each cilium we employ a semi-analytic method \cite{smith2009boundary}, utilising a constant-force approximation over each segment, which allows for analytical integration of the regularised-Stokeslet kernel.

The problem for one cilium, with time dependence omitted for brevity, is given by
\begin{equation}
  \begin{pmatrix}
    \mathcal{A}^{bb} & \, & \mathcal{A}^{bc}    \\[2em]
    \mathcal{A}^{cb} & \, & \mathcal{A}^{cc}    \\[0.2em]
  \end{pmatrix}
  \begin{pmatrix}
    \mathsf{F}^b     \\[2em]
    \mathsf{F}^c     \\[0.2em]
  \end{pmatrix}
  =
  \begin{pmatrix}
    \bm{0}     \\[2em]
    \dot{\mathsf{x}}^c     \\[0.2em]
  \end{pmatrix}
  ,\enspace
\mathcal{A}^{\sbt\mu} =
  \begin{pmatrix}
  	\, A_{11}^{\sbt\mu} &\, A_{12}^{\sbt\mu} &\, A_{13}^{\sbt\mu} \\[0.5em]
  	\, A_{21}^{\sbt\mu} &\, A_{22}^{\sbt\mu} &\, A_{23}^{\sbt\mu} \\[0.5em]
  	\, A_{31}^{\sbt\mu} &\, A_{32}^{\sbt\mu} &\, A_{33}^{\sbt\mu}
  \end{pmatrix}
  ,\enspace \mathsf{F}^{\sbt} =
  \begin{pmatrix}
  	\, \mathsf{F}_{1}^{\sbt\,}\\[0.5em]
  	\, \mathsf{F}_{2}^{\sbt\,}\\[0.5em]
  	\, \mathsf{F}_{3}^{\sbt\,}
  \end{pmatrix}
  ,\enspace \dot{\bm{\mathsf{x}}}^{c} =
  \begin{pmatrix}
  	\, \dot{\mathsf{x}}_{1}^{c}\\[0.5em]
  	\, \dot{\mathsf{x}}_{2}^{c}\\[0.5em]
  	\, \dot{\mathsf{x}}_{3}^{c}
  \end{pmatrix},
  \label{eq:blockSystem}
\end{equation}
where \(\lambda,\mu\in\{b,c\}\), and
\begin{equation}
    \begin{alignedat}{3}
    A_{ij}^{\sbt b}\{m,n\}   & = \frac{1}{8\pi}\sum_{q=1}^{Q^b} S_{ij}(\bm{x}^{\sbt\,}[m],\bm{X}^b[q]) \nu[q,n] \quad &\mbox{for}& \quad m=1,\ldots,N^{\sbt},\ n=1,\ldots,N^b , \\
    A_{ij}^{\sbt c}\{m,n\} 	 &= \frac{1}{8\pi}\int\limits_{s_n - \Delta s/2}^{s_n + \Delta s/2} S_{ij}(\bm{x}^{\sbt\,}[m],\bm{x}^c[n]) \mathrm{d}s\quad &\mbox{for}&\quad m = 1,\ldots, N^{\sbt},\ n=1,\ldots, N^c,
  \end{alignedat}
  \label{eq:blocks}
\end{equation}
where the integrals in \eqref{eq:blocks} are calculated analytically \cite{smith2009boundary}, with \(s_n = (n-1/2)\Delta s\) being the distance of point \(\bm{x}^c[n]\) along the cilium, and with the nearest-neighbour matrix \(\nu[q,n] = 1\) when \(n = \underset{\hat{n}=1,\ldots,N^\lambda}{\mbox{argmin}}\vert \bm{x}[\hat{n}] - \bm{X}[q]\vert\), and \(0\) otherwise. The problem \eqref{eq:blockSystem} can be initialised with the desired number of cilia by augmenting \(\mathsf{F}^c\) and \(\dot{\mathsf{x}}^c\) to take into account each cilia in order, for example, when modelling \(P\) cilia one would replace \(\dot{\mathsf{x}}_1^c\) with \(\{\dot{\mathsf{x}}_1^{c1},\ldots\dot{\mathsf{x}}_1^{cP}\}^T\).

For a collection of cilia rotating with the same fixed frequency, we can solve system \eqref{eq:blockSystem} over a single beat, and exploit the periodic nature of the problem by fitting a Fourier series to the calculated forces \(\mathsf{F} = \{\mathsf{F}^b,\mathsf{F}^c\}\), enabling ready estimation of these forces at any required time.

\begin{figure*}[tp]
    \centering
    \includegraphics[width=0.9\textwidth]{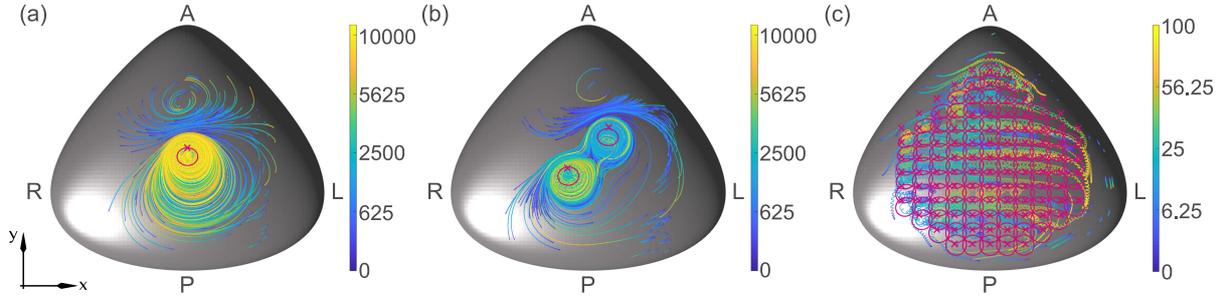}
    \caption{Simulated particle tracks for the (a) uniciliated, (b) biciliated, and (c) euciliated mouse node. Particles were released randomly in \(x,y\) at a height of \(0.2\) cilium lengths above the node base. Each track is coloured with respect to time (in cilia beats, detailed in the associated bars). In each panel the positions of the cilia are indicated with a cross, with the trace of the cilia tips shown in magenta.}
    \label{fig:tracks}
\end{figure*}

Solution of the resistance problem \eqref{eq:blockSystem} allows for calculation of the velocity field everywhere in the node, for example at a set of $N^p$ points \(\mathsf{x}^p = \{\mathsf{x}_1^p,\mathsf{x}_2^p,\mathsf{x}_3^p\}^T\), with corresponding fluid velocity at those points given by \(\mathsf{u}^p\), and evaluating the reduced motility problem for the velocity \({{\mathsf{u}}^p = \bra{\mathcal{A}^{pb}, \mathcal{A}^{pc}}\bra{\mathsf{F}^b,\mathsf{F}^c}^T}\). Extracellular particle tracking is carried out by the stochastic differential equation under advection and diffusion,
\begin{equation}
	\mathrm{d}\bm{x}^p = \bm{u}^p\mathrm{d}t + \mathrm{d}\bm{W}\quad\text{with}\quad\mathrm{d}\bm{W}\sim N_3\bra{0,2D\mathrm{d}t{\underline{\underline{I}}}\,},
	\label{eqn:update}
\end{equation}
Here the addition of Brownian noise is modelled using the Stokes-Einstein relation, with variance \(\sigma^2 = 2 D \mathrm{d}t\), and diffusion constant \(D = k_B T / 6\pi\eta a \approx 10^{-17}\), with \(k_B\) the Boltzmann constant, \(T\) temperature in Kelvin, \(\eta\) fluid dynamic viscosity, \(a\) the particle radius (taken to be \(50\)~nm before non-dimensionalisation), and \(N_3\left(0,\sigma^2{\underline{\underline{I}}}\,\right)\) being the 3-dimensional normal distribution with zero mean and variance \(\sigma^2\).

To model the oligociliated mouse node we take the geometric outline of Smith \textit{et al.} \cite{smith2011mathematical} improved with a more biologically realistic concave floor and a model of Reichert's membrane based on the microscopy images of Okada \textit{et al.} \cite{okada2005mechanism}, at its widest points the resulting node is approximately \(13.2\) cilium-lengths wide (R -- L), \(12\) cilium-lengths long (A -- P) and \(3.3\) cilium-lengths tall. Cilia are modelled as unit-length rigid rods with a semi-cone angle and posterior tilt of  \(30^\circ\). Numerical analysis of the convergence motivated the choice to model the geometry with \(2000\), \(4000\), and \(256\) vector degrees of freedom for stokeslet strength on Reichert's membrane, the nodal base, and rod-like cilia respectively, calculated over \(32\) time points for Fourier analysis. In the present work we investigate particle transport in two model systems, a uniciliated node with cilium at the centre of the node \({\bra{x,y} = \bra{0,0}}\), and a biciliated node with cilia offset at \(\bra{x_1,y_1}=\bra{-1,-1}\) and \(\bra{x_2,y_2}=\bra{1,1}\), where lengths have been scaled with respect to the length of a single cilium. The positions of the cilia in the biciliated node have been chosen as a representative system to test whether offset cilia, with separated cones of beating, in the centre of the node and thus away from extra boundary effects, can produce the long-range transport necessary for left/right symmetry breaking. As an initial investigation, in the present calculations we will consider each cilia to be beating with the same phase. For comparison between the mutant nodes and the wildtype, we also show calculations of a euciliated node with a more biologically realistic \(112\) cilia. Sketches for an individual beating cilium, and the geometry of the euciliated node are given in Figure~\ref{fig:sketch}.


\begin{figure*}[t]
    \centering
    \includegraphics[width=0.9\textwidth]{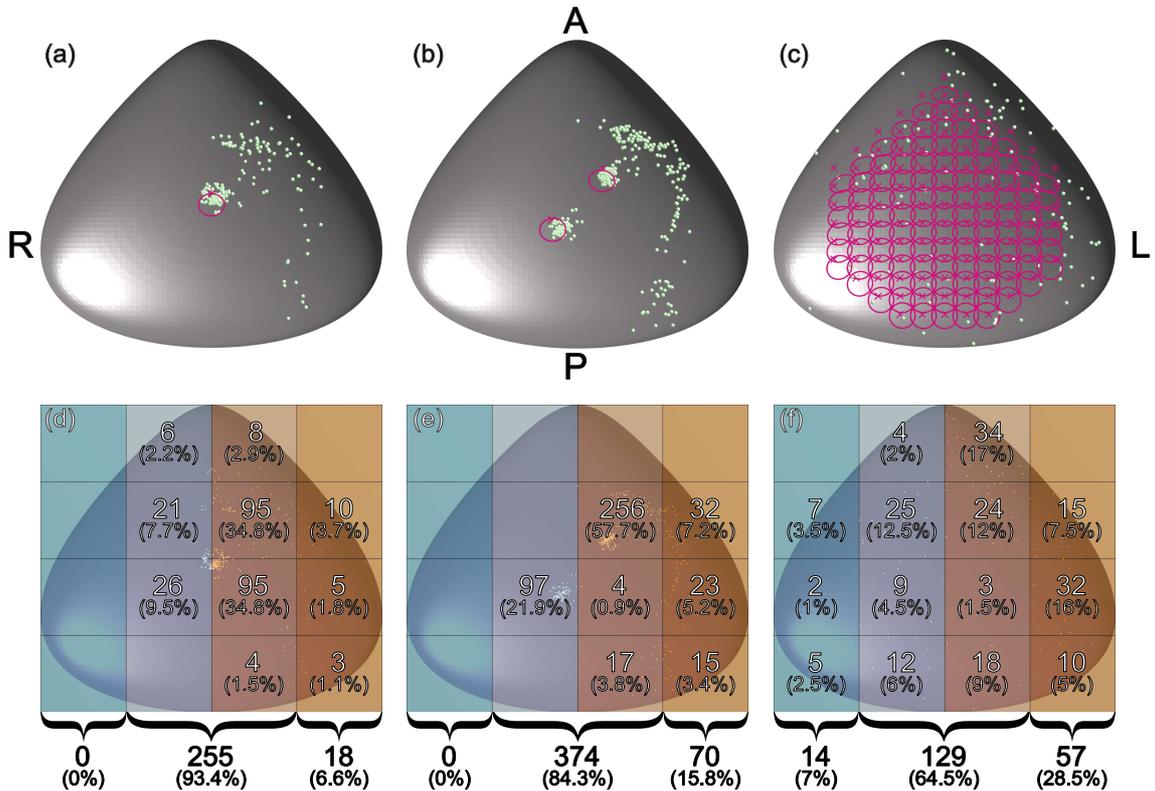}
    \caption{Deposition location of particles for the (a), (d) uniciliated, (b), (e) biciliated, and (c), (f) euciliated mouse node. Particles were released uniformly randomly in \(x,y\) at a height of \(0.2\) cilium lengths above the node base. Panels (a) -- (c) show particle deposition location as green balls, with cilia location marked by a cross, and path traced out by the cilia tips in magenta. Panels (d) -- (f) are divided into 12 regions with the number (and percentage) of particles deposed within these regions indicated. At the bottom of these panels the number (and percentage) of particles deposited in the far left, centre, and far right of the node is shown.}
    \label{fig:deposited}
\end{figure*}

\section{Transport of particles in the mouse node}

We investigate the potential for symmetry breaking via particle transport, paying particular attention to the difference in deposition in the case of the uniciliated node compared to that with flow driven by two beating cilia (which we call the biciliated node). Initially particles are released at a height of \(0.2\) cilium lengths above the node base, with position drawn uniformly in \(\bra{x,y}\) from within the node. In section~\ref{sec:2c_cr} an alternative release hypothesis that particles are released directly from the cilium themselves is considered. Particle locations are updated following equation~\eqref{eqn:update}, and tracked for \(10,000\) beats or, following \cite{omori2018simulation}, until they reach a height of \(0.1\) cilium lengths above the nodal floor, when they are considered deposited. The particle release assumption has been made to restrict the parameter space for this study, leaving more detailed consideration for future work. The computational cost of the simulations contained within this manuscript was just over \(3200\) cpu hours.

\subsection{Uniform release}

The euciliated node is initialised with \(200\) uniformly randomly released particles, all of which are deposited in fewer than \(100\) cilia beats. In contrast the uni- and biciliated nodes deposited approximately one third of the released particles in the \(10,000\) beat time period, with no significant difference in percentage of deposited particles in these cases (\(273\) deposited/\(946\) released in the uniciliated node, \(444\) deposited/\(1328\) released in the biciliated node). In each case the numbers of particles are sufficient to provide clear trends; the trends shown by the first half of the simulations are the same as those from the full dataset.

The particle tracks for each of the deposited particles are shown in figure~\ref{fig:tracks}. For each of the mutant mouse nodes we see a significant number of particles attracted towards the location of the cilia. Long-range transport of particles occurs in each case, with particles passing through a region directly anterior to the cilia being transported leftwards without being drawn into the deposition regions below the cilia. In contrast, the euciliated node (figure~\ref{fig:tracks}c) shows significant leftward transport of particles throughout the node. A selection of the individual paths can be seen in the supplementary material with figures S1 and S2 showing paths for the uniciliated node, S3 -- S9 the biciliated node, and S10 the euciliated node.

While the qualitative nature of transport in the uni- and biciliated nodes are similar, detailed analysis of the particle deposition location (figure~\ref{fig:deposited}) reveals that a particle is more than twice as likely to be deposited in the far-left region of the node when there are two motile cilia present than one, with \(7\%\) and \(16\%\) of particles reaching these regions for the uni- and biciliated nodes respectively). While in both cases this is markedly less than for the euciliated node (\(29\%\)), it highlights a significant difference between the mutants.

\subsection{Near-cilium wall release}

Solowiej-Wedderburn \textit{et. al} \cite{solowiej2019} hypothesised that particle exocytosis may be enhanced by wall stress. Based on the finding that the wall stress is highest in the cells adjacent and posterior to the sweeping cilium, they argued that localised particle release may combine with cilia-driven flow to effect left-right asymmetry in particle concentrations. By considering the tracks of particles that initiate in this region we can now see that the region of fast flow local to the beating cilia is not the primary location for long-range leftwards particle transport in the oligociliated node. While particles in the near-cilium region have high velocity, if we focus our attention only to particles released close to the cilia in figure~\ref{fig:tracks} we see that these paths are restricted in extent, ensuring local deposition rather than significant leftward bias. Instead, the primary region of long-range leftwards transport is in the lower-stress region anterior to the posterior-tilted cilia. This effect occurs because the inner field of the beating cilium is rotational, whereas the outer field is directional [30], an effect which can be explained by considering the image system of a rotlet near a plane boundary [12].


\begin{figure*}[tp]
    \centering
    \includegraphics[width=0.9\textwidth]{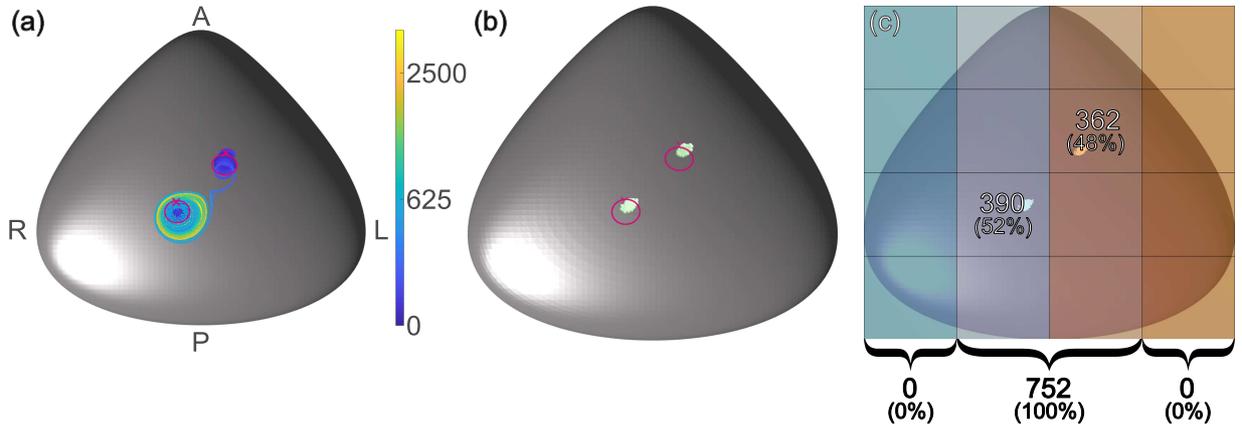}
    \caption{Results for particles released directly from the cilia of the biciliated mouse node. Panel (a) shows the simulated tracks, coloured according to time in beats with the cilia location marked with a cross and path traced out by the cilia tips shown in magenta. In panel (b) the deposition location of the particles is shown, with panel (c) showing the node divided into 12 regions with the number (and percentage) of particles deposed within these regions indicated. At the bottom of panel (c) the number (and percentage) of particles deposited in the far left, centre, and far right of the node is shown.}
    \label{fig:2c_cr}
\end{figure*}

\subsection{Cilium release}
\label{sec:2c_cr}

Another potential mechanism, suggested during discussions at the Theo Murphy International Meeting associated with this volume, is that the morphagen-laden particles may be released directly from the cilium itself. To test this hypothesis we release particles at a randomly chosen distance along the cilium \(0.2 < s < 1\) (where \(s\) is measured in cilium lengths) and at a random time in the beat cycle and solve for the particle trajectories, as before, until they are designated deposited or until \(10,000\) beats have elapsed. Figure~\ref{fig:2c_cr} depicts the paths traced out by the 752 deposited particles (\(\approx 79\%\) of the \(952\) initialised particles), together with the deposition locations of each particle. While these figures show that some particles can be passed from a region close to the most leftward cilium to the most rightward, no extended transport is seen in the leftwards direction and all particle deposition is confined to the central ciliated regions (\(48\%\) and \(52\%\) deposited near the leftward and rightward cilia respectively), thereby having no significant left-right bias.

\section{Discussion}

In this manuscript we have addressed the problem of symmetry breaking in the mouse embryonic node through numerical simulation of the flow driven by motile cilia in the uni- and biciliated mutant, as well as a \(112\) cilium wildtype.  We have shown that, if the signal-carrying particles are released uniformly throughout the node, then a single beating cilium is sufficient to effect some leftward bias in particle transport and deposition, with two beating cilia producing approximately double the amount of bias.

Two models for localised particle release were considered. The first, motivated by recent work \cite{solowiej2019}, examined the idea of particle release from the high stress region localised to the wall region posterior to the beating cilia. The second, suggested in discussions at the Theo Murphy International meeting of March 2019, was that particles may be released from the motile cilia themselves. Despite the intuitive appeal of these models, neither produced significant particle transport in the oligociliated node, and hence do not appear to be consistent with experimental observations in mouse \cite{shinohara2012two}.

In the present work we have not explicitly considered the presence of immotile cilia, thought by some to be essential for detecting particles \cite{yoshiba2012}. However, our abstraction of particle deposition based on reaching a threshold height above the node floor leaves room for interpretation of the mechanism of deposition. Provided there is a good covering of immotile cilia for detection across the base of the node this is unlikely to change the findings of the present work, although a further detailed study with individually modelled immotile cilia could assess this impact.

These observations lead us to hypothesise that, in the mouse at least, a morphogen-laden particle transport model may be feasible, with spatially-extended particle release being a more probable scenario than cilium-localised release. The situation may of course be different in other species, particularly in the other key experimental system of zebrafish Kupffer's vesicle, in which much greater ciliary activity is necessary for consistent \emph{situs} (see \cite{amack2004,okabe2008,wang2012,gokey2016,kim2017}). The computational tools we describe in this paper, which enable flexibility over node morphology, cilia location, tilt, and activity, should, with minor modification, be widely applicable to assess particle transport across many of the diverse node structures that have been reported, including rabbit, chick, and medaka.

There is still a large morphology and parameter space to explore within the mouse system, taking into account variability in cilia position, length, relative phase, tilt angle and beat frequency, and indeed the detailed microarchitecture of the ciliated epithelium - significant further computational study to assess these features is required. Other localised release mechanisms may also be possible. For example if the posterior-tilted cilium exerts an elastic stress on the convex cell from which it protrudes, this may induce localised particle release in the directional flow region anterior to the cilium. We cannot therefore draw sweeping conclusions from the simulation studies presented here, but rather we offer these findings as part of the continuing dialogue between experiment and theory, and between molecular developmental biology and mechanics.

\section{Acknowledgements}

The authors would like to thank the organisers and participants of the Theo Murphy International Meeting associated with this volume, particularly for the insightful discussions that helped formulate some of the theories tested in this manuscript. We would like to thank Susana Lopes and her group at the Nova University of Lisbon for helpful discussions and on-going collaborations surrounding the symmetry-breaking problem in zebrafish. Finally, the authors would like to acknowledge the late John Blake who was fundamental in initiating and inspiring our work in this field.

\section{Funding}

MTG and DJS gratefully acknowledge support from the Engineering and Physical Sciences Research Council (EPSRC) Award No. EP/N021096/1. TDMJ gratefully acknowledges support from EPSRC Award No. EP/R041555/1.

\section{Author contributions}

MTG, DJS and TDMJ designed the research; MTG conducted simulations and analysed results; MTG, DJS and TDMJ co-wrote the paper.

\section{Competing intersets}

We have no competing interests.

\end{document}